# ROOM TEMPERATURE ON-WAFER BALLISTIC GRAPHENE FIELD-EFFECT-TRANSISTOR WITH OBLIQUE DOUBLE-GATE

Mircea Dragoman[1*], Adrian Dinescu[1], and Daniela Dragoman[2,3]

[1]National Institute for Research and Development in Microtechnology (IMT), P.O. Box 38-160, 023573 Bucharest, Romania,

[2]Univ. Bucharest, Physics Faculty, P.O. Box MG-11, 077125 Bucharest, Romania

[3]Academy of Romanian Scientists, Splaiul Independentei 54, 050094 Bucharest, Romania

## Abstract

We have fabricated and measured ballistic graphene transistors with two oblique gates that can be independently biased. The gate lengths are about 38 nm and are separated by a distance of 30 nm, the tilting angle being of $45^{o}$ with respect to source and drain electrodes distanced at 190 nm. Electric measurements reveal specific properties of ballistic carrier transport, i.e. nonlinear drain voltage-drain current dependence, showing a saturation region, and negative differential resistance at certain bias voltages, which cannot be explained without physical mechanisms related to ballistic transport. Tens of ballistic transistors, with very large transconductances, were fabricated on a chip cut from a 4 inch graphene wafer. Such double-gate transistor configurations can be used also as extremely efficient, state-of-the-art photodetectors.

**1. Introduction**

Ballistic transistors are searched for THz cutoff frequencies and as ultrafast devices for logic operations. The main goal is room temperature operation for ballistic field-effect-transistors (FETs), but this aim can be accomplished only if the transistor channel is made from materials displaying large mean-free paths of carriers at room temperature. For example, in carbon nanotube transistors the mean-free-path is greater than 1 μm at room temperature [1,2], while in the ballistic deflection transistor, which has a THz cutoff frequency and was fabricated by etching its six-terminal coplanar structure in InGaAs, this parameter attains 200 nm at room temperature [3].

High-quality graphene monolayers display one of the largest mean-free-paths, of around 400 nm if graphene is deposited on $SiO_2$, or up to 1 μm in graphene deposited over boron nitride [4]. Very recently, it was found that the room-temperature mean-free path could be even longer, of more than 10 μm [5] in 20-100 nm wide graphene nanoribbons epitaxially grown on SiC. Moreover, extensive experimental data of a room-temperature graphene ballistic transistor matching simulations were recently reported [6], in a rather different configuration compared to a graphene CMOS-like transistor [7]. In addition, ballistic transport in a back-gate CMOS-like graphene transistor with a very short channel length (10-20 nm), in fact a nanogap channel created by applying rather strong DC electric fields to a constriction in the middle of a gold electrode [8], was recently evidenced. In the last case, the nanogap was produced by a break junction formed through electromigration on top of the graphene flake. Since the electric field is applied directly to graphene, the risk of damaging it is high, and thus the nanogap transistor cannot be easily reproduced and the fabrication has a low yield. Despite these limitations, the main imprint of ballistic transport is observed, i.e. nonlinear drain current-drain voltage dependence at various gate voltages due to variation of the mode density [8].

The aim of this paper is to demonstrate reproducible batch fabrication, with high yield, of room-temperature double-gate graphene ballistic transistors on a graphene wafer. The ballistic transistor in a double-gate configuration was theoretically studied in the context of THz generation [9-10] and reversible logic [11]. In the fabricated devices, the two gates are obliquely placed at an angle of $45^o$, the small dimensions of the transistor guaranteeing ballistic transport regime, in particular the appearance of transmission bandgaps, without significant degradation of the physical properties of the graphene monolayer [12-14]. In addition, we show that the double-gated graphene transistors act also as efficient photodetectors.

Ballistic transport described by the Landauer equation, together with the possibility of opening a bandgap in transmission due to the oblique-gate configuration creates very strong nonlinear drain current-drain voltage dependences at various gate voltages, and even negative differential resistance regions. Therefore, we anticipate a rather high transconductance and saturation regions, characteristics that are opposite to non-ballistic graphene FETs, where the same dependences are linear or almost linear. The configuration with double oblique gates is thus expected to surpass the properties of common graphene FETs configurations, involving normally placed gate electrodes.

## 2. Fabrication

The CVD grown graphene used in this experiment was transferred on a silicon dioxide layer (300 nm thickness) thermally grown on a high resistive silicon wafer. The graphene transfer was performed by Graphene Industries. After the graphene transfer, a Raman analysis on the entire wafer revealed that about 80% of the wafer contains high quality graphene monolayer. We have considered the graphene monolayer as defectless if the Raman spectrum contains a peak in the G band at around 1590 cm$^{-1}$ and the 2D peak located around 2645 cm$^{-1}$, if the ratio between the 2D and G peaks is higher than 1.8 and if no defect band D appears in the Raman signal. Figures 1(a) and 1(b) illustrate Raman spectra of high-quality and poor-quality graphene monolayer, respectively.

Further, we have cut a chip from the wafer in a region with high-quality graphene monolayer, and fabricated on it the double-gated graphene ballistic transistor depicted in Fig. 2. The two tilted gates (G1 and G2) are separated from the graphene and the drain (D) and source (S) electrodes by a layer of HSQ (hydrogen silsesquioxane) dielectric with a thickness of 40 nm measured using ellipsometry. The fabrication process involves twelve different steps, mainly: electron beam lithography (EBL), reactive ion etching (RIE), metal deposition, and lift-off. The graphene ribbons (the channels of the future FETs) were patterned by EBL using a dedicated equipment: RAITH e-Line, and then etched by RIE in oxygen plasma (equipment: SENTECH Etchlab 200) and precisely placed in a set of previously fabricated register marks (see Figs. 3(a) and 3(b)). Working with a very thin (50 nm) PMMA (poly(methyl methacrylate)) layer, the S and D contacts were patterned and then deposited by electron beam evaporation of Ti/Au (7 nm/30 nm), as illustrated in Figs. 3(c) and 3(d). The small thickness of both the metal layer and the electron resist are required by the particular geometry of the electrodes and by the small distance between them, of 190 nm. In order to fabricate the gates, a very thin negative electron resist (HSQ-XR1541-2) was deposited by spinning, and then shaped by EBL, forming a 40 nm thick insulating layer on top of the S and D contacts, as shown in Fig.

3(e). The two gate electrodes, tilted at $45^{o}$ with respect to these electrodes, were first patterned in 70 nm thick PMMA and then fabricated by lift-off from Ti (7 nm) and Au (40 nm), as can be seen from Fig. 3(f). The gate lengths are about 38 nm and the distance between them is of 30 nm. The contact pads were patterned also by EBL. 10 nm of Ti and 100 nm of Au were deposited by highly directional electron beam evaporation equipment (TEMESCAL FC-2000) in order to achieve a good quality of the lift-off process. In Fig. 4(a) we present the SEM image of the area of the two tilted gates, the optical image of the transistor being shown in the inset.

We have fabricated in the batch 70 transistors, with the same configuration as in Fig. 4(a), and 20 transistors with coplanar (CPW) lines as electrodes, for microwave applications; the first (second) type of transistors will be referred to in the following as FETs with thin (CPW) electrodes. A CPW electrode consists of three electrodes, the central one for signals and other two, outer electrodes for grounds. All these electrodes (see Fig. 4(b)) are much larger than those illustrated in Fig. 4(a) and were fabricated on top of the first electrodes, by the same metallization technique. The CPW electrodes are Ti/Au (10 nm/300 nm).

## 3. Measurements and discussions

We have performed on-wafer measurements of all transistors using Keithley 4200 SCS equipment with low noise amplifiers at outputs. The DC probes connected to the Keithley 4200 are positioned, together with the entire probe station, in a Faraday cage. The chip containing the graphene FETs is placed on the chuck of the probe station, the DC probes are connected to the three electrodes of the graphene FET, and then the Faraday cage is closed. The measurements are done via a computer, which monitors the entire system. All measurements are performed at room temperature by grounding the source electrode. In order to verify the validity of our measurements, we have repeated the experiments at different voltage steps. Intentionally, we did not introduce any smoothing procedures during the measurements or afterwards. The measurements revealed that 23% of all transistors do not work at all, due to various reasons, especially due to defects in graphene and/or exfoliated metallizations caused by lack of adherence. The rest of the transistors are working and have similar behaviours. The graphene chip containing tens of ballistic FETs positioned on the chuck inside the probe station and with DC probes positioned on a transistor is illustrated in Fig. 5.

In Fig. 6(a), we have displayed a typical drain current ($I_D$) versus drain voltage ($V_D$) dependence of the graphene ballistic FET with CPW electrodes, when both gates are biased at the same voltage $V_{G1} = V_{G2} = V_G$, because in FETs with coplanar electrodes measured with probe tips only two bias voltages are allowed, i.e. a drain voltage and a gate voltage. The values of the common gate voltages are indicated in the inset of Fig. 6(a). From this figure, it can be seen that at small gate voltages the $I_D$-$V_D$ dependence is strongly nonlinear, in deep contrast with non-ballistic graphene transistors where the same dependences are linear. The tendency to saturation is evident beyond a drain voltage of 1.7 V. Although, for clarity reasons, we have represented $I_D$-$V_D$ for positive drain voltages only, the transport of carriers is ambipolar and a similar $I_D$-$V_D$ dependence is obtained for negative drain voltages. The nonlinear $I_D$-$V_D$ dependence, in particular the saturation tendency in these curves, is a result of the Landauer formula valid in the ballistic transport regime. The appearance of a negative differential resistance region at high positive gate voltages (at 25 V in Fig. 6(a)) is caused by the opening of a gap in the transmission coefficient between source and drain, gap that is a signature of the ballistic transport regime across oblique gates (see Refs. [6] and [13]).

In Fig. 6(b) we have depicted the dependence of $I_D$ on the gate voltage $V_G$ at several values of the drain voltage, indicated in the inset. We see that, at a constant gate voltage, the drain current increases with $V_D$. The minimum value of $I_D$ as a function of $V_G$ for $V_D = 0.5$ V is associated to the change in shape of the $I_D$-$V_D$ curves in Fig. 6(a) due to the appearance of the transmission gap at high positive gate voltages.

In Fig. 7(a) we have displayed the drain conductance dependence on $V_D$, dependence that is symmetric at negative and low positive gate voltages. The values of the drain conductance in these cases are higher than the transconductance values, as given in Fig. 7(b), which indicates that there is no gain in these regions. This is a common situation in many graphene FETs where the intrinsic gain is small or even absent. However, at high positive gate voltages, where the drain conductance becomes asymmetric with respect to $V_D$ and decreases dramatically, becoming even negative at drain voltages corresponding to the negative differential conductance observed in Fig. 6(a), there is a narrow region around $V_D = 0.7$ V where the drain conductance is smaller than the transconductance. In this region, the transistor has gain/is amplifying. As follows from Fig. 7(b), the transconductance has a relative maximum value at a drain voltage of -2 V and a gate voltage of 8 V.

The transistors with thin electrodes, as those in Fig. 4(a), were measured by independently biasing the two oblique gates. Nonlinear $I_D$-$V_D$ dependences at different gate voltages, with saturation tendencies and negative differential resistance regions have been observed also in these situations, a typical example being presented in Fig. 8. In this case one gate was biased with a constant voltage of $V_{G2}$ = -3 V, while the other gate bias was varied at the values indicated in the inset. A first observation is that the drain current is much higher compared to graphene FETs with CPW electrodes (see Fig. 6(a)), and that $I_D$ varies significantly even at very low gate voltages. In particular, a negative differential resistance region is observed at a value of only $V_{G1}$ = 1 V. The higher drain currents in graphene FETs with thin electrodes compared to FETs with CPW electrodes can be explained by a lower contact resistance in the former case compared to the latter.

The drain conductance of the transistor in Fig. 8, depicted in Fig. 9(a) for positive $V_D$ values, has the same behaviour and comparable values as in graphene FETs with CPW electrodes. In particular, it becomes negative for $V_{G1}$ values for which a gap in the transmission coefficient/a negative differential resistance region occurs. The transconductance of the same transistor with very thin electrodes, represented in Fig. 9(b) as a function of $V_{G1}$ for different $V_D$ values indicated in the inset, is one order of magnitude higher than in graphene CPW FETs (see Fig. 7(b)). Moreover, these high $g_m$ values are obtained at very low gate voltages. This result is to be expected because a high contact resistance (as in FETs with CPW electrodes) has an exponentially detrimental impact on the peak transconductance [15]. The maximum transconductance value, of 50 μS at a drain voltage of 3 V and a gate voltage of -0.5 V, corresponds to 1.42 mS/μm when scaled to the gate length, which is a very good, state-of-the-art value (see the maximum transconductance table collected from different scientific works in [16]). In addition, in a wide $V_D$ range, between 1.7 V and 2.5 V, the drain conductance is lower than the transconductance, the graphene FET showing an important gain.

Encouraged by the high currents and large sensitivity to gate voltages of the double-gated ballistic graphene FET, we investigated also the photoresponse of this device by illuminating it with a red laser diode having a power of $P = 0.5$ mW. The photoresponse, i.e. the difference $\Delta I = I - I_{dark}$ between the current measured in the presence and absence of illumination, is represented in Fig. 10 as a function of $V_D$, for several $V_{G1} = V_{G2} = V_G$ gate voltages given in the inset. Assuming that the diameter of the laser diode is about 1.5 mm, and knowing that the dimensions of the graphene flake are $0.37 \times 1.7$ $\mu m^2$, the responsivity has a value of $R = 2.8 \times 10^5$ A/W for $\Delta I = 50$ $\mu A$. The responsivity was calculated as $R = \Delta I / P_{eff}$, where $P_{eff} = (A_{flake} / A_{laser})P$, with $A_{flake}$ and $A_{laser}$ the area of the graphene flake and of the laser beam, respectively. This responsivity value is quite impressive, and could not be attained for photodetectors based on graphene monolayers in any other configuration. Indeed, efficient graphene-based photodetectors usually require heterostructures with other two-dimensional materials with bandgap, or band structure engineering of graphene in order to introduce a bandgap. Otherwise, the responsivities of graphene-based photodetectors, as well as of those fabricated from other two-dimensional materials have, in general, smaller values than $10^5$ A/W at room temperature (see [17]). For example, photodetectors based on graphene quantum-dot like arrays and defect midgap states show responsivities of 8.61 A/W [18], while hybrid graphene quantum dots can attain responsivities as high as $10^8$ A/W, and in graphene-MoS$_2$ heterostructures $R$ is as high as $10^7$ A/W [19]. Therefore, the responsivity of the double-gated ballistic graphene FET reported in this paper is state-of-the-art, this result being achieved not by introducing bandgaps in graphene or by using heterostructures, but by inducing bandgaps in the ballistic transmission coefficient of graphene via appropriately chosen biases applied on the gates. Because the bandstructure of graphene monolayer is not modified in this device configuration, the photoresponse is expected to have a large bandwidth.

## 4. Conclusions

We have fabricated room-temperature ballistic graphene FETs with two oblique gates that can be independently biased on a wafer scale. The ballistic nature of carrier transport is demonstrated by the nonlinear electrical characteristics and even by the appearance of negative differential regions for some gate voltages, phenomena associated to ballistic propagation of charge carriers in graphene in the presence of oblique gates. The nonlinear $I_D$-$V_D$ characteristics, which tend to saturate at quite low drain voltages, are accompanied by large transconductances, especially in devices with thin, independently biased electrodes, which make them extremely interesting for nanoelectronics. In addition, these ballistic graphene-based FETs show state-of-the-art responsivities as photodetectors, without any modification to the unique, bandgapless electronic structure of monolayer graphene. The specific configuration of the device, involving oblique gates, takes advantage of the particularities of Dirac electron propagation, being precisely tailored for electronic and optoelectric applications of this material.

*Acknowledgements* We acknowledge the financial support of Nucleus Project TEHNOSPEC PN 1632 2015-2017.

**Figure captions**

Fig. 1  Raman spectra of (a) high-quality and (b) poor-quality monolayer graphene.

Fig. 2  SEM imagine of the ballistic graphene transistor.

Fig. 3  (a) Graphene ribbon placed between the alignement marks, (b) optical image of the graphene channel, (c) optical image of S and D electrodes, (d) high magnification SEM micrograph of the electrodes, (e) optical microscope image of the HSQ patterned area acting as  the gate dielectric, and (f) optical microscope image of the double gate tilted electrodes.

Fig. 4  (a) SEM image of the graphene transistor with this electrodes in the gate area, with the optical image of the transistor in inset, and (b) SEM image of the CPW ballistic transistor.

Fig. 5  The graphene ballistic FET under measurement in the probe station with the DC probes positioned on transistor electrodes.

Fig. 6  (a) The $I_D$-$V_D$ dependence of the graphene FET with CPW electrodes at different gate voltage values (the same on both gates) given in the inset, and (b) the $I_D$-$V_G$ dependence of the graphene FET with CPW electrodes (the same $V_G$ is applied on both electrodes) at different drain voltage values given in the inset.

Fig. 7  (a) The dependence of the drain conductance of the graphene FET on drain voltage at various gate voltages (the same on both gates) indicated in the inset, and (b) the dependence of the transconductance of the graphene FET on gate voltage (the same on both gates) at different drain voltages indicated in the inset.

Fig. 8   The $I_D$-$V_D$ dependence of the graphene ballistic FET with thin electrodes at various gate voltages applied on $G_1$ indicated in the inset (the voltage on $G_2$ is -3 V in all cases).

Fig. 9 (a) The drain conductance dependence on drain voltage at several gate voltages on $G_1$ indicated in the inset, and (b) the transconductance dependence on gate voltage on $G_1$ at various drain voltages given in the inset. The voltage on $G_2$ is -3 V in all cases.

Fig. 10 Dependence of photocurrent on drain voltage at several gate voltages (the same on both gates) indicated in the inset.

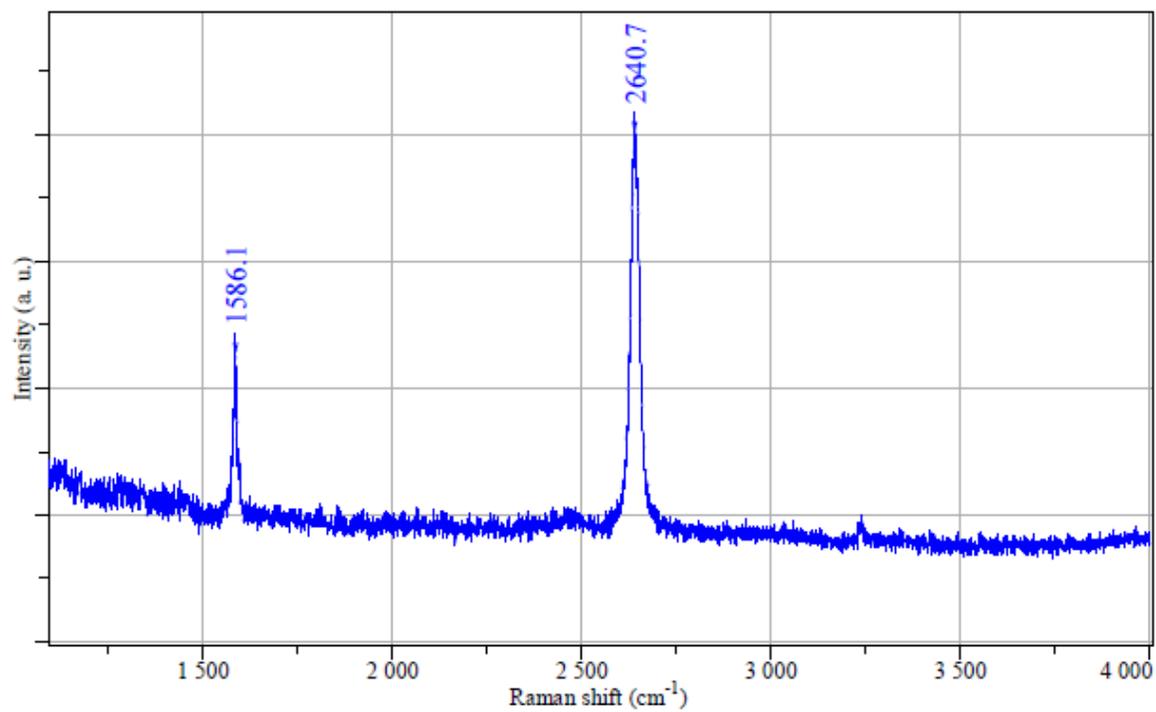

(a)

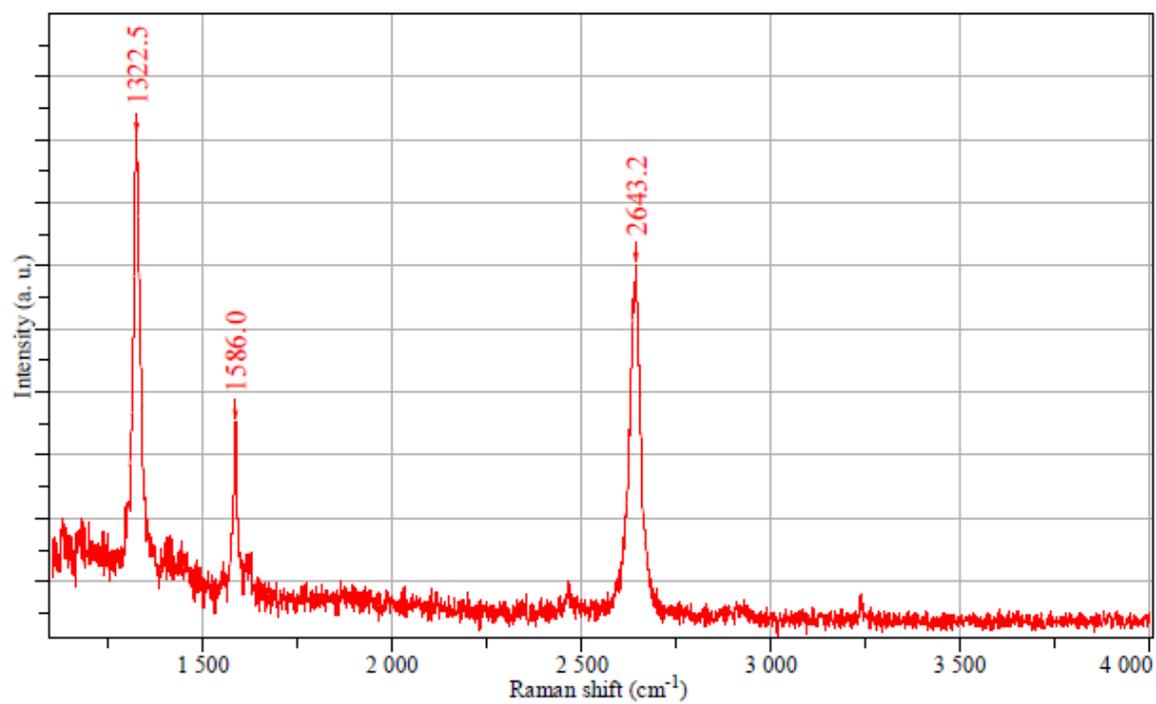

(b)

Fig. 1

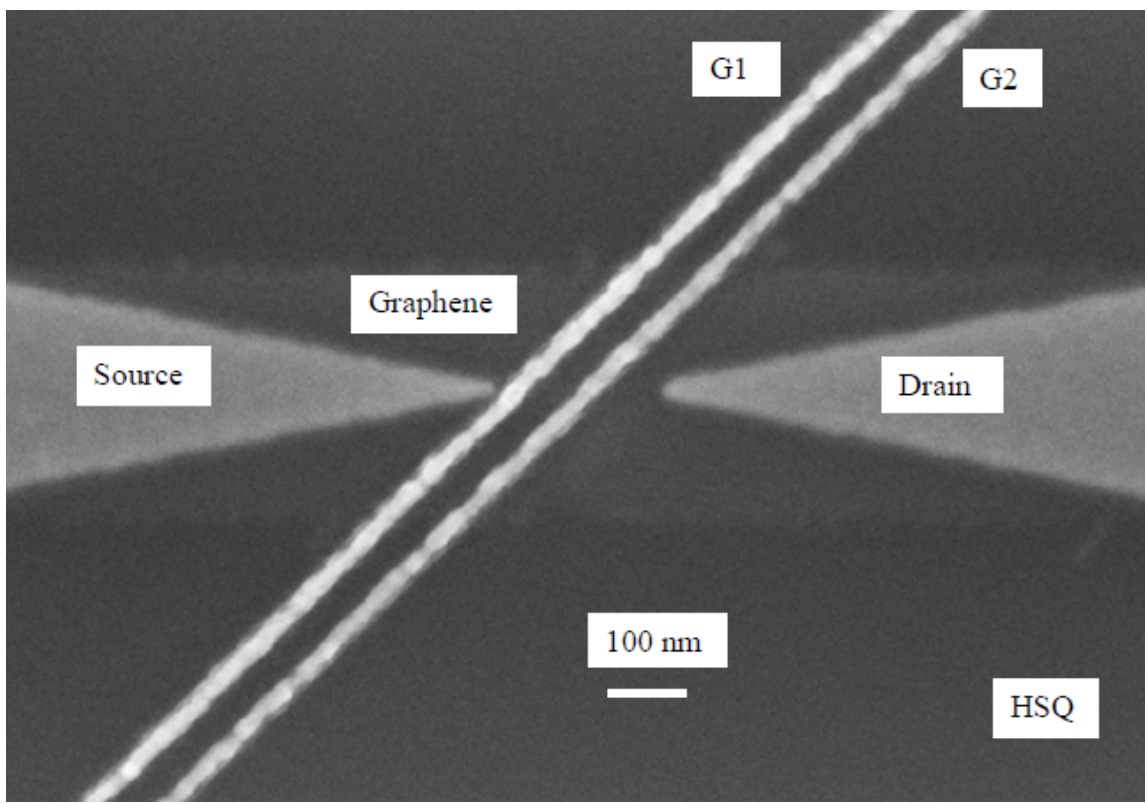

Fig. 2

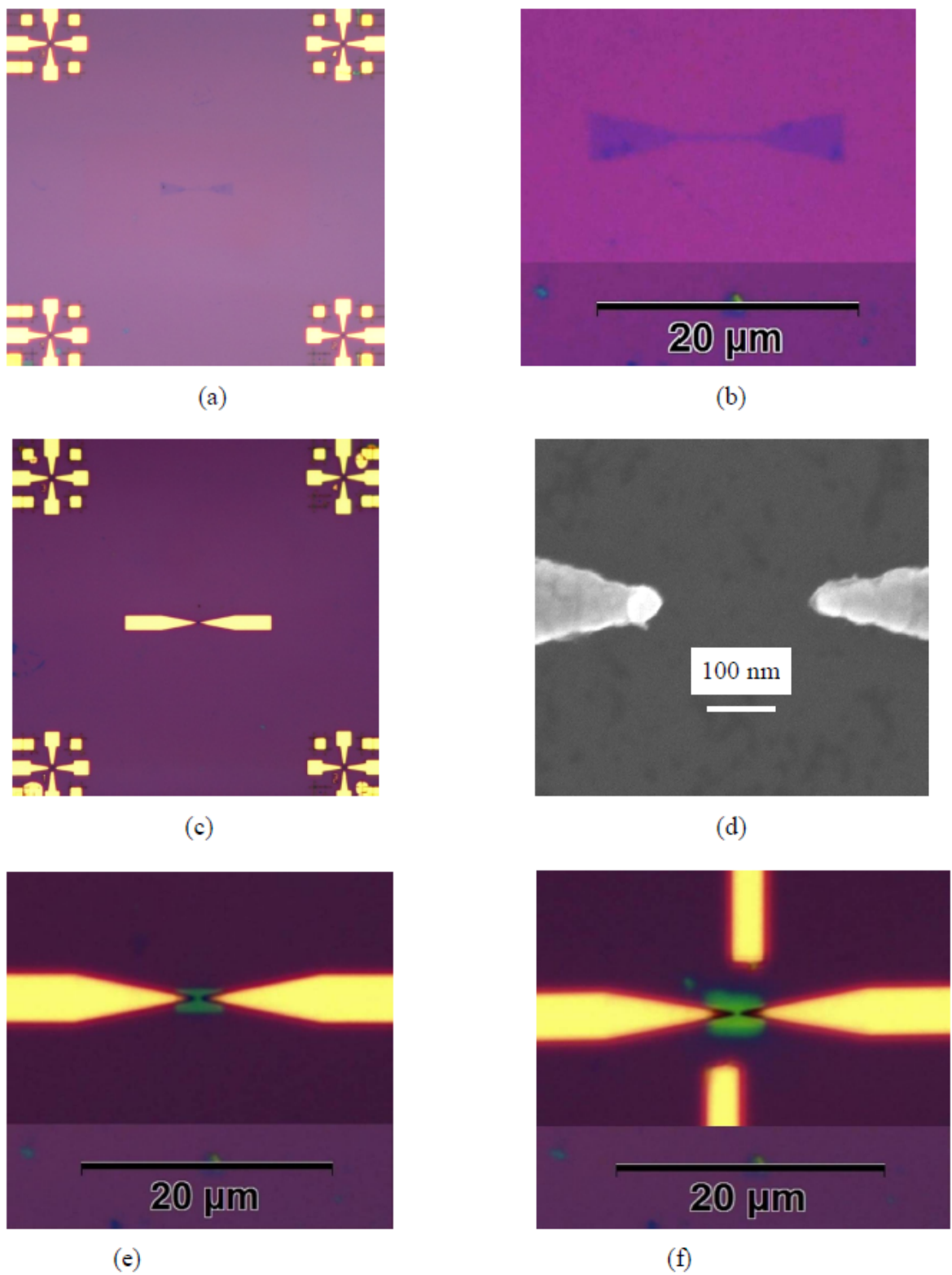

(a)

(b)

20 µm

(c)

(d)

100 nm

(e)

20 µm

(f)

20 µm

Fig. 3

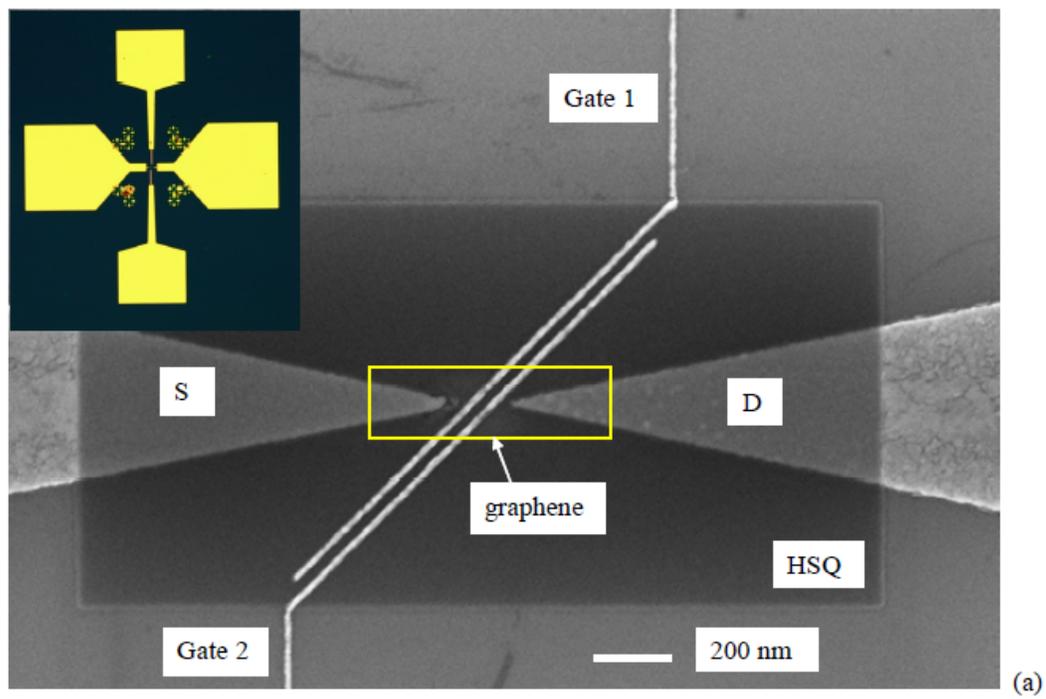

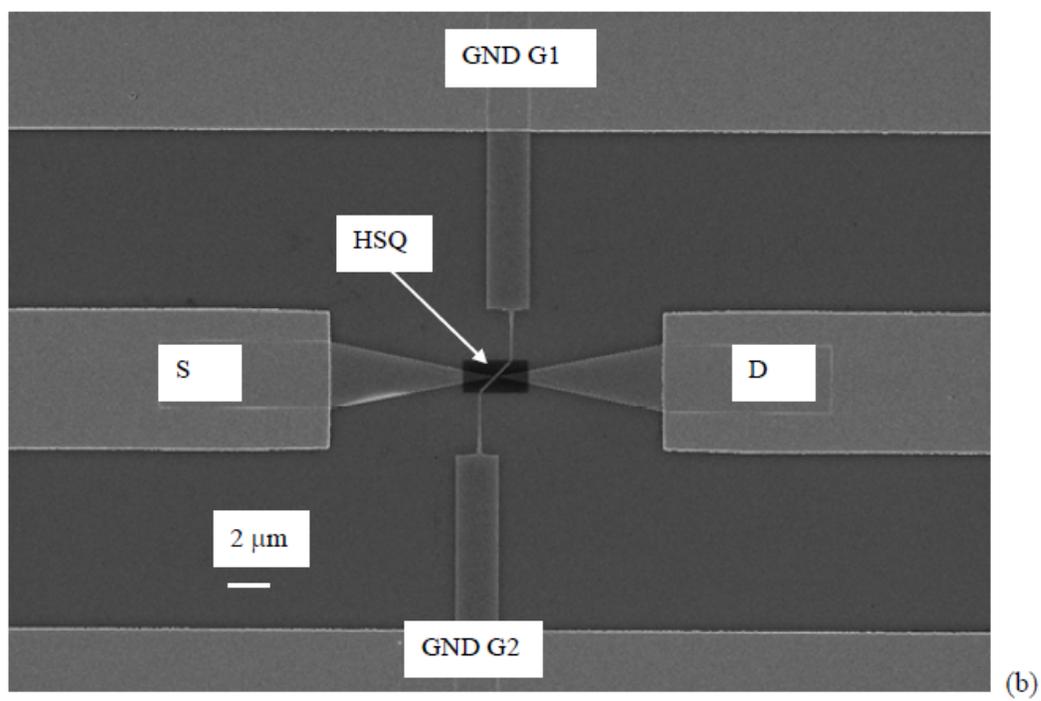

Fig. 4

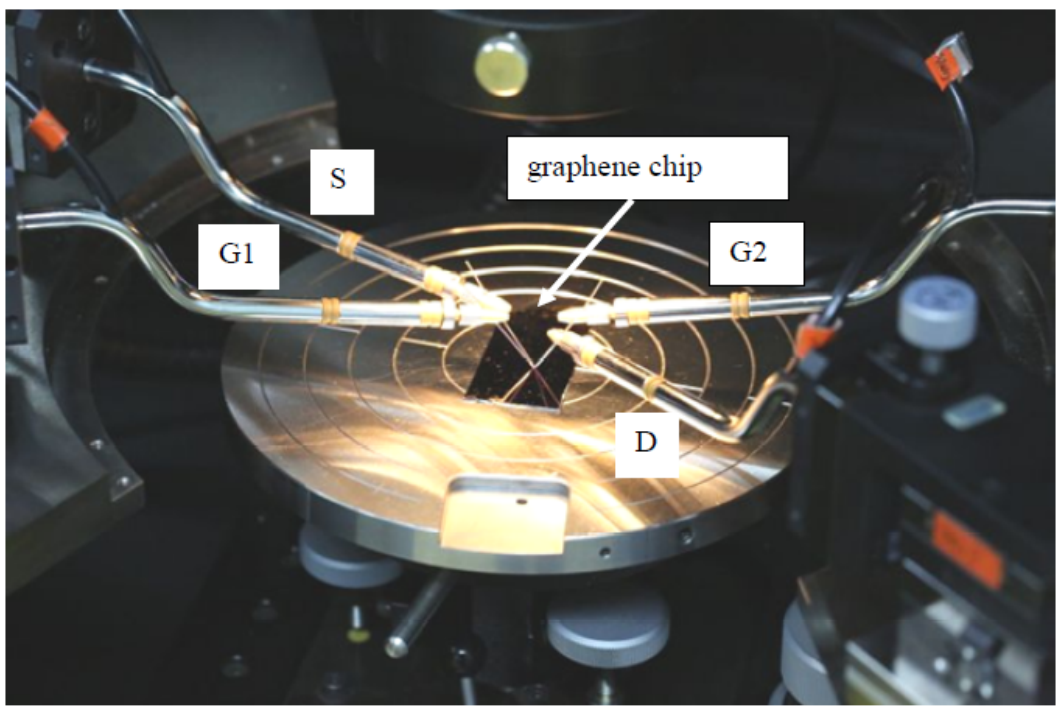

Fig. 5

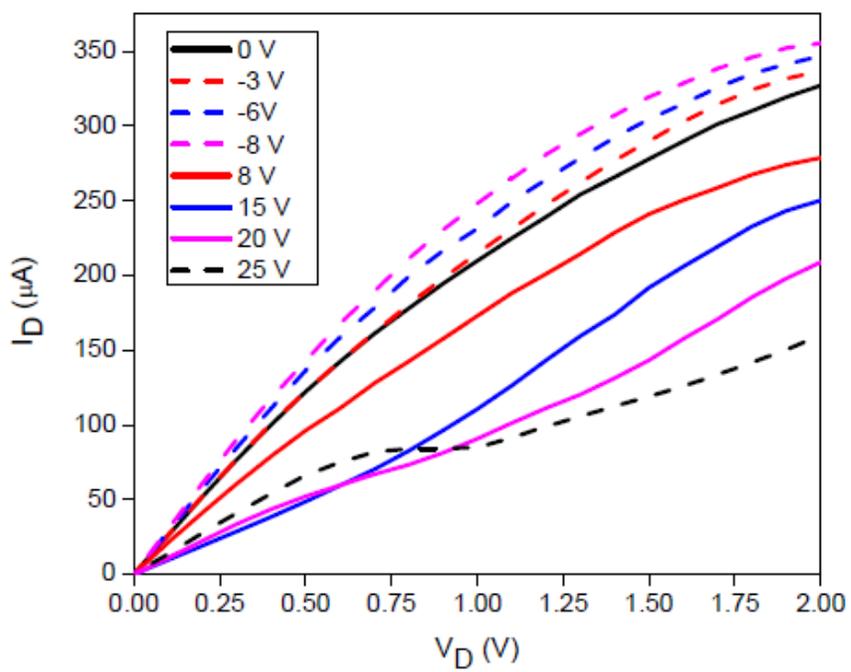

(a)

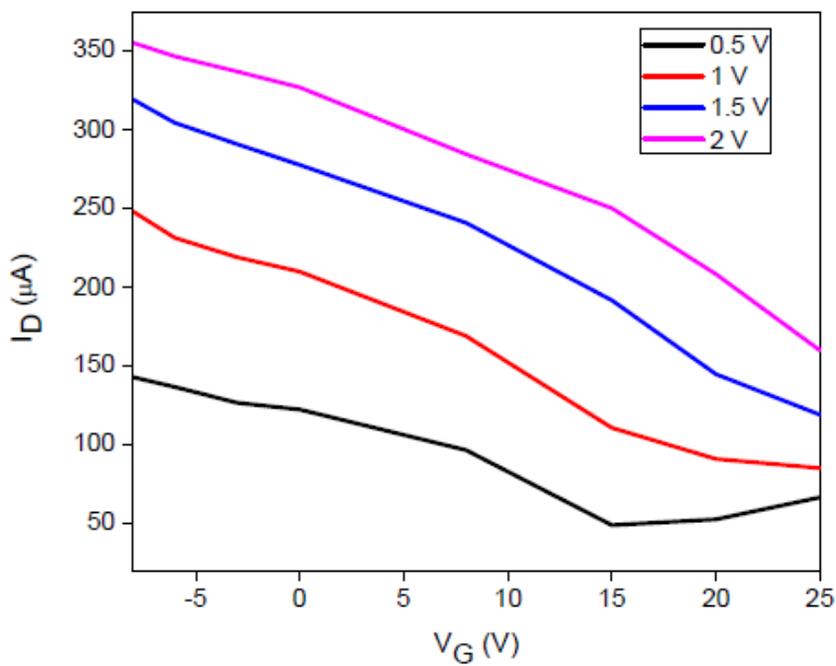

(b)

Fig. 6

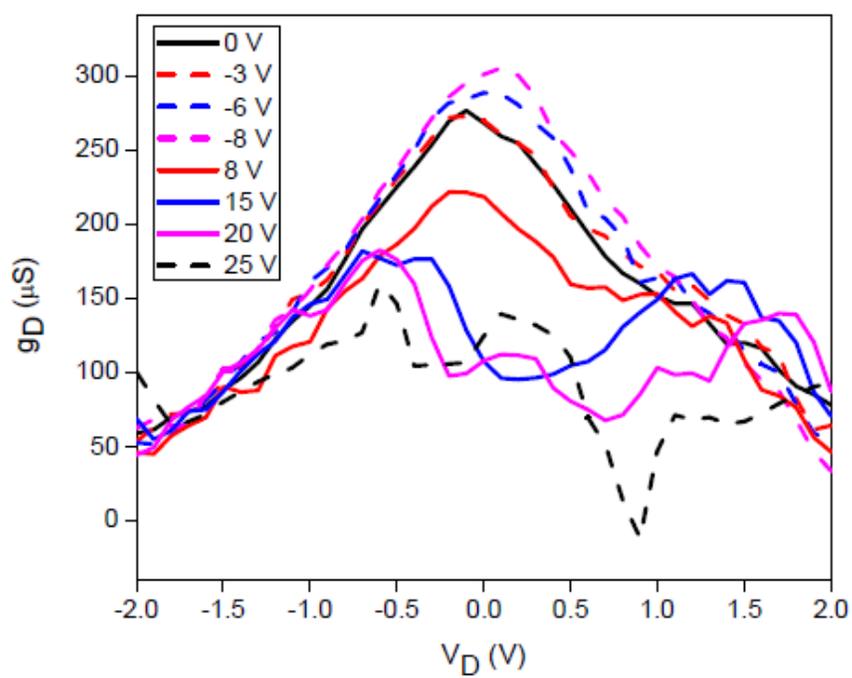

(a)

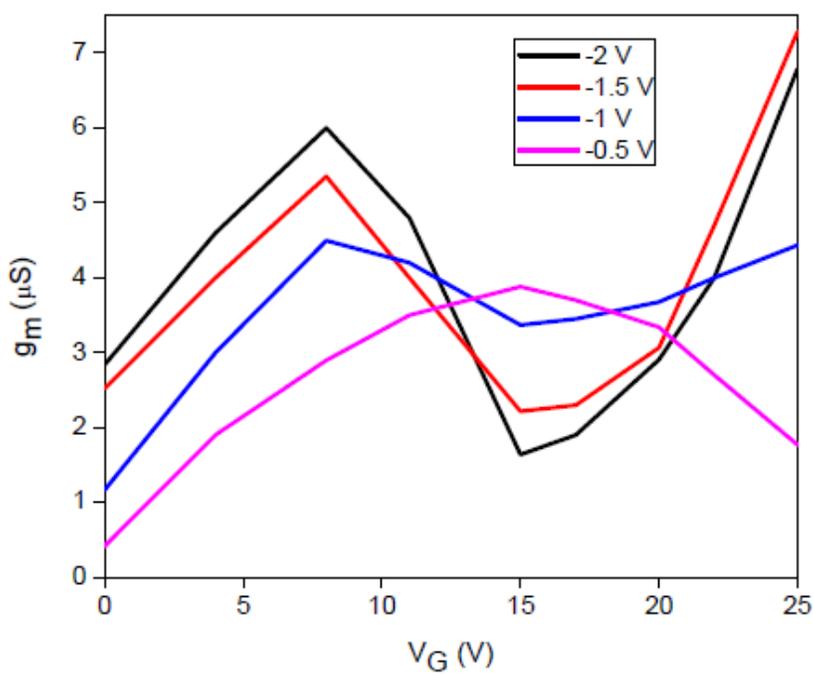

(b)

Fig.7

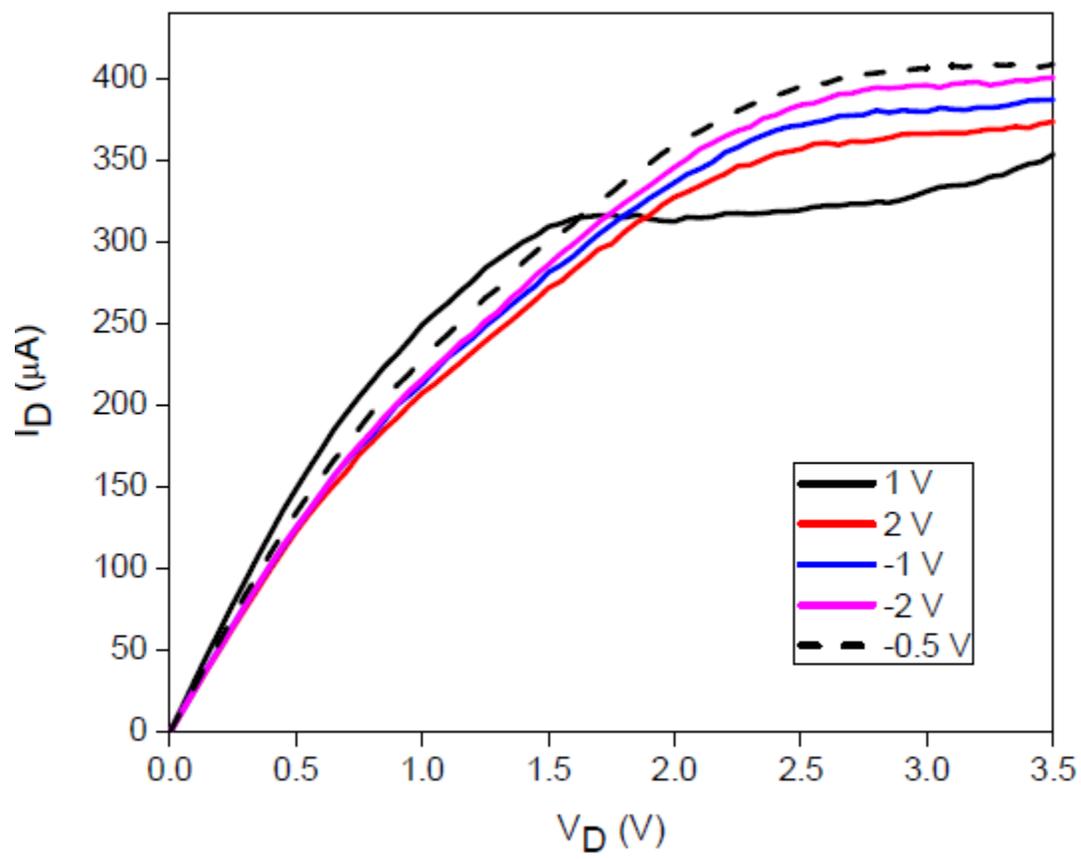

Fig. 8

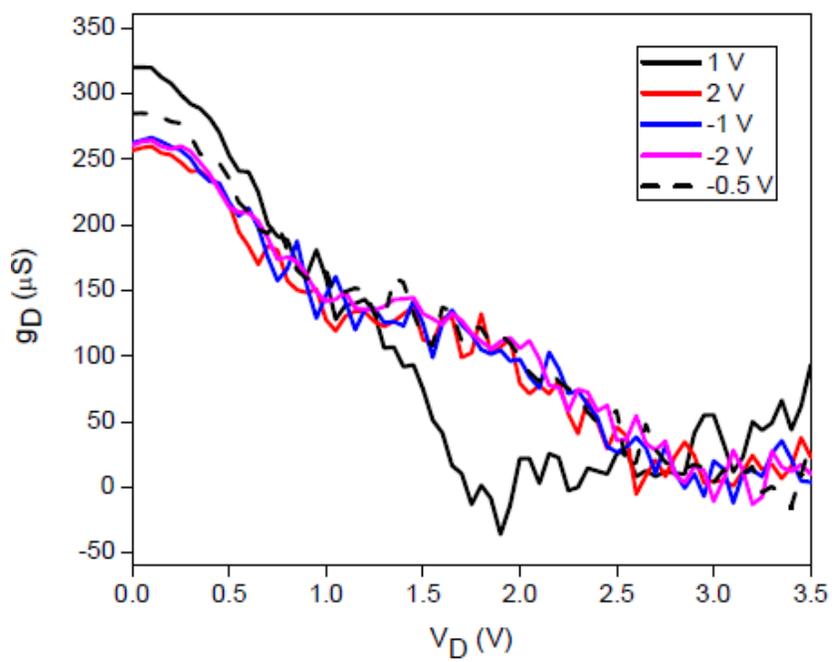

(a)

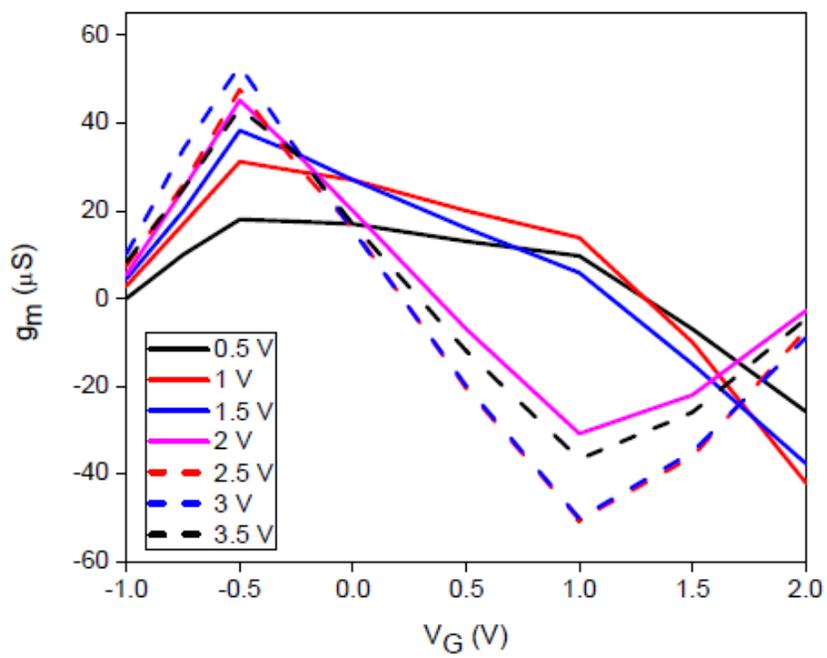

(b)

Fig. 9

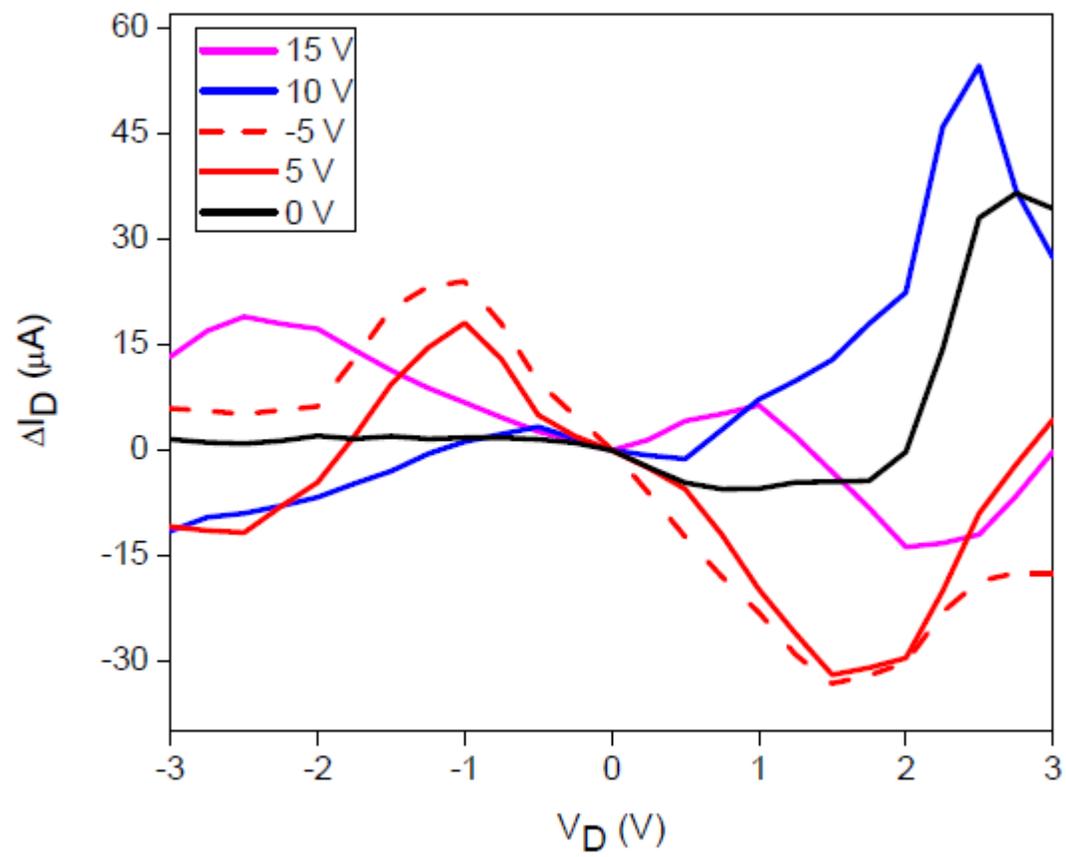

Fig. 10